\documentclass[twocolumn]{aastex63}

\received{\today}
\revised{}
\accepted{}
\submitjournal{ApJ}

\shorttitle{Dropouts of fully stripped ions}
\shortauthors{Raymond et al.}
\graphicspath{{./}{figures/}}

\begin{document}

\def\RSUN{R$_{\sun}$}
\def\kms{$\rm km~s^{-1}$}
\def\cmcube{$\rm cm^{-3$}$}
\def\O3{[O~III]}
\def\Hetwo{He$^{2+}$}
\def\Cfive{C$^{5+}$}
\def\Csix{C$^{6+}$}
\def\Nseven{N$^{7+}$}
\def\Oeight{O$^{8+}$}
\def\Oseven{O$^{7+}$}
\def\Osix{O$^{6+}$}
\def\Ofive{O$^{5+}$}

\title{Dropouts of Fully Stripped Ions in the Solar Wind: A Diagnostic for Wave Heating versus Reconnection}

\correspondingauthor{John C. Raymond}
\email{jraymond@cfa.harvard.edu}

\author[0000-0002-7868-1622]{John C. Raymond}
\affiliation{Center for Astrophysics $|$ Harvard \& Smithsonian,
60 Garden St.,
Cambridge, MA 02138, USA}

\author{M. Asgari-Targhi}
\affiliation{Center for Astrophysics $|$ Harvard \& Smithsonian,
60 Garden St.,
Cambridge, MA 02138, USA}

\author{Maurice L. Wilson}
\affiliation{Center for Astrophysics $|$ Harvard \& Smithsonian,
60 Garden St.,
Cambridge, MA 02138, USA}

\author{Yeimy J. Rivera}
\affiliation{Center for Astrophysics $|$ Harvard \& Smithsonian,
60 Garden St.,
Cambridge, MA 02138, USA}

\author{Susan T. Lepri}
\affiliation{Department of Climate and Space Sciences and Engineering, University of Michigan, Ann Arbor, MI 48109, USA}

\author{Chengcai Shen}
\affiliation{Center for Astrophysics $|$ Harvard \& Smithsonian,
60 Garden St.,
Cambridge, MA 02138, USA}




\begin{abstract}
The SWICS instrument aboard the ACE satellite has detected frequent intervals in the slow solar wind and interplanetary coronal mass ejections (ICMEs) in which \Csix\/ and other fully stripped ions are strongly depleted, though the ionization states of elements such as Si and Fe indicate that those ions should be present.  It has been suggested that these ``outlier" or ``dropout" events can be explained by the resonant cyclotron heating process, because these ions all have the same cyclotron frequency as \Hetwo.  We investigate the region in the corona where these outlier events form.  It must be above the ionization freeze-in height and the transition to collisionless plasma conditions, but low enough that the wind still feels the effects of solar gravity.  We suggest that the dropout events correspond to relatively dense blobs of gas in which the heating is reduced because local variations in the Alfv\'{e}n speed change the reflection of Alfv\'{e}n waves and the turbulent cascade.  As a result, the wave power at the cyclotron frequency of the fully stripped ions is absorbed by \Hetwo\/ and may not be able to heat the other fully-stripped ions enough to overcome solar gravity.  If this picture is borne out, it may help to discriminate between resonant cyclotron heating and stochastic heating models of the solar wind.

\end{abstract}




\section{INTRODUCTION} \label{sec:intro}

Models for heating and driving the solar wind generally emphasize either dissipation of magnetohydrodynamic (MHD) wave energy or magnetic reconnection.  The wave models can naturally explain strong preferential heating of O and Mg ions observed in the fast solar wind and at large heights in the slow wind \citep{kohl97, cranmer99, frazin03}. The reconnection models can naturally explain the observed density fluctuations and some composition anomalies in the slow wind.  Both models have some difficulty reproducing the ionization states measured in the heliosphere \citep{oran15, shen17, szente22}.  We also note that while coronal mass ejections (CMEs) are basically driven by simple MHD forces, the ejected plasma continues to be heated after it leaves the Sun \citep{akmal01, lee09, rakowski07, rakowski11, murphy11, wilson22}, but the nature of that heating is difficult to determine.  

Composition anomalies in the solar wind provide a means to connect observed structures in the wind to their coronal origins, and they constrain the physical processes that drive the wind.  The first composition changes to be measured were variations in the He abundance correlated with solar wind speed \citep[][]{ogilvie1974, kasper12, alterman19}.  It was later established that abundances of elements whose First Ionization Potential (FIP) was below about 10 eV were enhanced in the slow solar wind \citep{geiss1995, vonsteiger2000}.  More recently, drastic dropouts in the abundances of heavy elements, probably associated with clouds of depleted gas released from the cusps of streamers, were reported \citep{weberg12, weberg15}. 

Two recent papers have reported remarkable ionization state anomalies, which they call ``Outliers", in the solar wind.  \citet{zhao17} examined data from the Solar Wind Ion Composition Spectrometer (SWICS) instrument aboard the Advanced Composition Explorer (ACE) satellite, and they found intervals when the ratio \Csix / \Cfive\/ was anomalously low by as much as an order of magnitude compared with other measures of the ionization state, such as \Oseven / \Osix. \citet{zhao17} examined the slow solar wind, and \citet{kocher17}, performed a similar study of interplanetary CMEs (ICMEs).  Almost half of the anomalies occurred in the slow wind and almost half in ICMEs, with  only about 1\% of the events occurring in the fast wind. More recently, \cite{rivera21} extended that work to include magnetic field measurements and more elemental composition data.  Notably, \citet{rivera21} show changes in the magnetic field direction and the electron strahl properties of the Outlier wind indicating a strong connection with heliospheric current sheet crossings. 

\citet{zhao17} and \citet{kocher17} list several other properties of the outlier intervals.  In both the slow solar wind and ICMEs, the fluxes of N$^{7+}$ and O$^{8+}$, that is to say the ions that are stripped bare, are anomalously low, along with the \Csix.  The ionization states of other elements, such as Ne, Si, S and Fe, and their FIP fractionation seem to be the same as that of the slow solar wind or CME in which the outliers are embedded.  In the slow wind, the outlier intervals occur about 10\% of the time, and the densities during the outlier intervals are about twice normal.  The proton temperature is somewhat lower than is typical for the slow wind, and perhaps bimodal.  Under normal circumstances, much of carbon would be in the \Csix\/ state.  The total abundance of carbon is depleted, unlike the other elements.  Thus it seems that ions in the stripped ionization state are simply missing, rather than being shifted to lower ionization states. On the other hand, Rivera et al. (2021) find that the He/H ratios are enhanced  relative to regions upstream and dowstream of the outliers, and they discuss magnetic field variations associated with the dropouts.

Here, we try to quantify the requirements on the fractionation mechanism.  We explore one mechanism, a lack of wave energy at the resonant frequency of the fully stripped ions, in more detail.  The large abundance of \Hetwo\/ means that \Hetwo\/ can absorb all the available wave energy at that frequency if the heating rate is sufficiently low.  In many models for the origin of the solar wind, Alfv\'{e}n waves from the solar surface partially reflect from the Alfv\'{e}n speed gradient in the corona, and the interaction between inward- and outward-propagating waves generates a turbulent cascade that can heat the ions.   Density fluctuations in the corona cause variations in the Alfv\'{e}n speed and the wave reflection, so they create significant variation in the wave local heating rate. Some models for the observed preferential heating of ions such as O VI and Mg X are based upon resonant absorption of wave power at the cyclotron frequency \citep{cranmer00, hollweg02}.  However, it is problematic whether the turbulence can cascade to the resonant frequency.  Therefore, other models rely on stochastic heating by lower-frequency Kinetic Alfven Waves at the scale of the ion gyroradius \citep{chandran10}.  The outlier events discussed here may offer a means to discriminate between these two pictures.

From here on, we will take the term ``Alfv\'{e}nic waves" to mean Alfv\'{e}n-like waves including fast mode waves and Kinetic Alfv\'{e}n Waves (KAWs). The paper is organized as follows: Section 2 reviews the observed characteristics of the Outlier wind, Section 3 summarizes the physical mechanisms that have been discussed to explain them, and Section 4 lists constraints on the region where the Outlier wind must be formed.  In Section 5 we consider how wave-driven and reconnection-driven winds might produce the dropouts, and in Section 6 we discuss the different implications of dropouts in the slow solar wind and in ICMEs.  Section 7 summarizes our conclusions.

\section{GENERAL CHARACTERISTICS} \label{sec:general}

We first summarize the observed characteristics of the dropout regions as determined by \citet{zhao17} for the slow wind, \citet{kocher17} for ICMEs, and \citet{rivera21} for both the slow wind and ICMEs.

In the slow solar wind, the dropouts show depletions of \Csix\/ by up to an order of magnitude.  The wind speeds of 350 to 450 \kms and the \Oseven/\Osix\/ ratios and Fe charge states are normal for the slow wind.  The dropouts occur about 10\% of the time, and they show density enhancements by factors of 1.2 to 2, with proton temperatures reduced by about 50\%. Helium abundances are also enhanced by factors of 1.2 to 2, and the FIP bias is normal for the slow wind, except that Ne  in the ICME outliers is enhanced relative to nearby regions in the ICMEs.  The magnetic field shows rotations or sudden changes in direction, and enhanced magnetic field fluctuations are seen downstream of the dropouts. Alfv\'{e}n speeds are enhanced by a factor of 1.7 on average.  Unidirectional electron strahl is observed over half of the time.

In ICMEs, reductions of \Nseven\/ and \Oeight\/ are seen, though otherwise the ionization state is similar to that of the surrounding plasma.  Dropouts are seen in 72\% of the ICMEs studied by \citet{rivera21}.  The FIP enhancements are typical of ICMEs, but Ne is enhanced, as in the slow solar wind.

\citet{rivera21} suggest a link between the dropouts seen in the slow wind and in ICMEs, in that the magnetic field changes seen in slow wind dropouts are similar to what is seen in the small interplanetary magnetic flux ropes (SIMFRs) studied by \citet{moldwin00}, \citet{feng08}, \citet{kilpua09} and \citet{huang20}.  In particular, they show strong rotation of the magnetic field, and they are often associated with sector boundaries, elevated He abundance, elevated $\rm <Q_{Fe}>$ and electron strahl.  These point to an association with reconnection events in the corona.

An important consideration is that the fractionation would not occur in a steady flow.  If the particle fluxes are set by the boundary condition at the chromosphere and the plasma is in a steady flow, then the flux of each species must remain constant.  In principle, there could be variations between one streamline and its neighbor.  High density contrasts among magnetic flux tubes over scales of a few thousand km were inferred from eclipse images \citep{november96} and from Atmospheric Imaging Assembly (AIA) observations of Comet C/2011 (Lovejoy) by \citet{raymond14}.  Thus the relatively steady flow in the fast wind may be part of the explanation for the lack of dropouts in the fast wind.

Assuming that the outliers develop just above the freeze-in and collisional/collisionless transition, we take 4 \RSUN\/ as the typical location.  If an Outlier lasts for two hours at a solar wind speed of 400 \kms, the size scale is around $3 \times 10^{10}$ cm, which corresponds to about $6 \times 10^8$ cm at 4 \RSUN.  If the speed increases from 100 \kms at 4 \RSUN\/ to 400 \kms at 1 AU and the area increases as r$^2$, a density of 10 $\rm cm^{-3}$ at 1 AU would correspond to a density of $10^5$ $\rm cm^{-3}$ at 4 \RSUN. 

Several characteristics of the source regions of the slow solar wind and CMEs are likely to be relevant to the origin of the dropouts.  In the case of the slow wind, the acceleration is slow, reconnection is likely to be  important, and the heliospheric current sheet is present. The wind is unsteady and clumpy in comparison with the fast solar wind, though the coronal holes where the fast wind originates also show modest density inhomogenities \citep{cadavid19, hahn18}.  Both coronal holes and quiet Sun regions have low-lying magnetic loops, but only the quiet Sun and active regions have high coronal loops.  In the case of  ICMEs, the range of characteristics makes it difficult of point to the important characteristics.  They have complex structures, including a leading edge, a flux rope (void), a prominence(core), and an EUV dimming region.  There can be a broad range of speeds and densities, and reconnection plays an important role. 

\section{PHYSICAL CAUSES OF DROPOUTS}

\citet{zhao17} consider several possible interpretations for the anomalous \Csix\/ dropouts in the slow solar wind.  They raise the possibilities that 1) resonant heating by MHD waves favors ions that are not fully stripped, 2) closed magnetic loops depleted in fully stripped ions reconnect with open field lines to release their plasma, or 3) fully stripped ions are depleted by Coulomb collisions with energetic protons produced by magnetic reconnection.  

Scenario 2, the opening of loops depleted in fully stripped ions, is attractive because it is related to models for the FIP (First Ionization Potential) effect \citep{schwadron99, laming17, laming19}, and it fits in with the intermittent appearance of the anomalies in the slow solar wind.   However, the abundance of S suggests that solar wind FIP fractionation does not happen on the same closed loops responsible for the bulk of the coronal EUV and X-ray emission,
but that magnetic structures without strong resonances like open field regions may play a role (J.M. Laming 2022, private communication).  Moreover, attributing the dropouts to opening loops just shifts the problem from open to closed field regions without offering a specific mechanism.  Also, it is not obvious how many of the loop openings would occur above the freeze-in height.  Scenario 3 posits a source of energetic protons. \citet{simnett95} emphasized the role of energetic protons in flares, and proton beams could be produced during reconnection between open and closed magnetic field believed to be important in causing the FIP effect in the slow solar wind \citep{schwadron99, laming17}. However, this scenario has the difficulty that the cross-section for momentum transfer between protons and \Csix\/ is only about 44\% larger than that for \Cfive, so that \Csix\/ would not be much more strongly depleted than \Cfive.  Moreover, sufficiently energetic protons would ionize \Cfive, producing an anomaly opposite to that observed.  

It therefore seems that the first scenario is the most likely because, as pointed out by \citet{zhao17}, all the missing ions have the same charge-to-mass ratio and therefore share the same cyclotron frequency.   Evidence for the importance of preferential heating  in driving the heavy element component of the solar wind comes from the large velocity widths of O and Mg ions observed in coronal holes and in streamers above 3 \RSUN\/ \citep{kohl97, frazin03}, though the measured line profiles are not good enough to distinguish between resonant and stochastic heating \citep{klein16}.  Large variations in the $\rm ^3He / ^4He$ ratio observed across the solar cycle \citep{gloeckler16} may be compatible with either picture.  If there is a deficit of wave energy at that frequency, all the fully stripped ions will experience less resonant heating.  It is also conceivable that there is excess energy at the resonant frequency of the fully stripped ions, so that they are heated and driven out of a parcel of plasma.  However, they would have to be spread over a large volume to avoid producing regions with excess fully stripped ions, and no such regions are reported by \citet{zhao17} or \citet{rivera21}. 

\citet{kocher17} measured the ionization state anomaly in ICMEs.  Much less is known about the heating of CME plasma than about the heating of streamers.  It is generally assumed that the plasma in CMEs is ejected by purely MHD forces, which would not naturally produce the fractionation observed.  However, observations of CMEs with UVCS \citep{kohl95, kohl97} show that the plasma continues to be heated well after it is launched \citep{akmal01, ciaravella01,lee09,murphy11, wilson22}, and the ionization states of CME plasma at 1 AU with ACE and STEREO also require heating at least to heights of several \RSUN\/ \citep{rakowski07, rakowski11, rivera19}. The nature of that heating is not fully settled, but dissipation of magnetic energy, presumably via turbulence, is the most viable option \citep{kumar96, landi10, murphy11}.  If that turbulence cascades to small enough scales, it could resonantly heat the ions.

\section{LOCATION OF DROPOUT FORMATION}

\subsection{Ionization Freeze-in}

As pointed out by \citet{zhao17}, the fractionation must occur above the point where ionization states freeze-in.  Otherwise, the \Cfive\/ would have time to ionize to \Csix\/ before the plasma leaves the low corona.  The freeze-in height depends on the density, temperature and velocity of the wind, so it varies greatly among different coronal features.  It also depends upon the ions being considered, since the ionization and recombination rates of ions such as \Cfive\/ and \Csix\/ are relatively slow compared with the Si and Fe ions with L-shell or M-shell electrons at coronal temperatures.  Therefore, the carbon ionization state will freeze in first, at lower heights.  The freeze-in occurs very low in the fast solar wind because of its high speed and low density.  We are interested in the slow wind, where freeze-in typically occurs between about 2 \RSUN\/ and 3 \RSUN\/ in the pseudostreamer modeled by \cite{shen17}, though it varies within the pseudostreamer structure as the wind velocity increases and the density decreases away from the neutral line.  \citet{landi14} found that the usual freeze-in picture cannot account for the full range of ionization states in the fast solar wind, possibly due to non-Maxwellian electron distributions \citep{ko96}, but perhaps due to the breakdown of the assumption of steady flow in the models.   

\begin{figure*}
\begin{centering}
\includegraphics[width=5.0in]{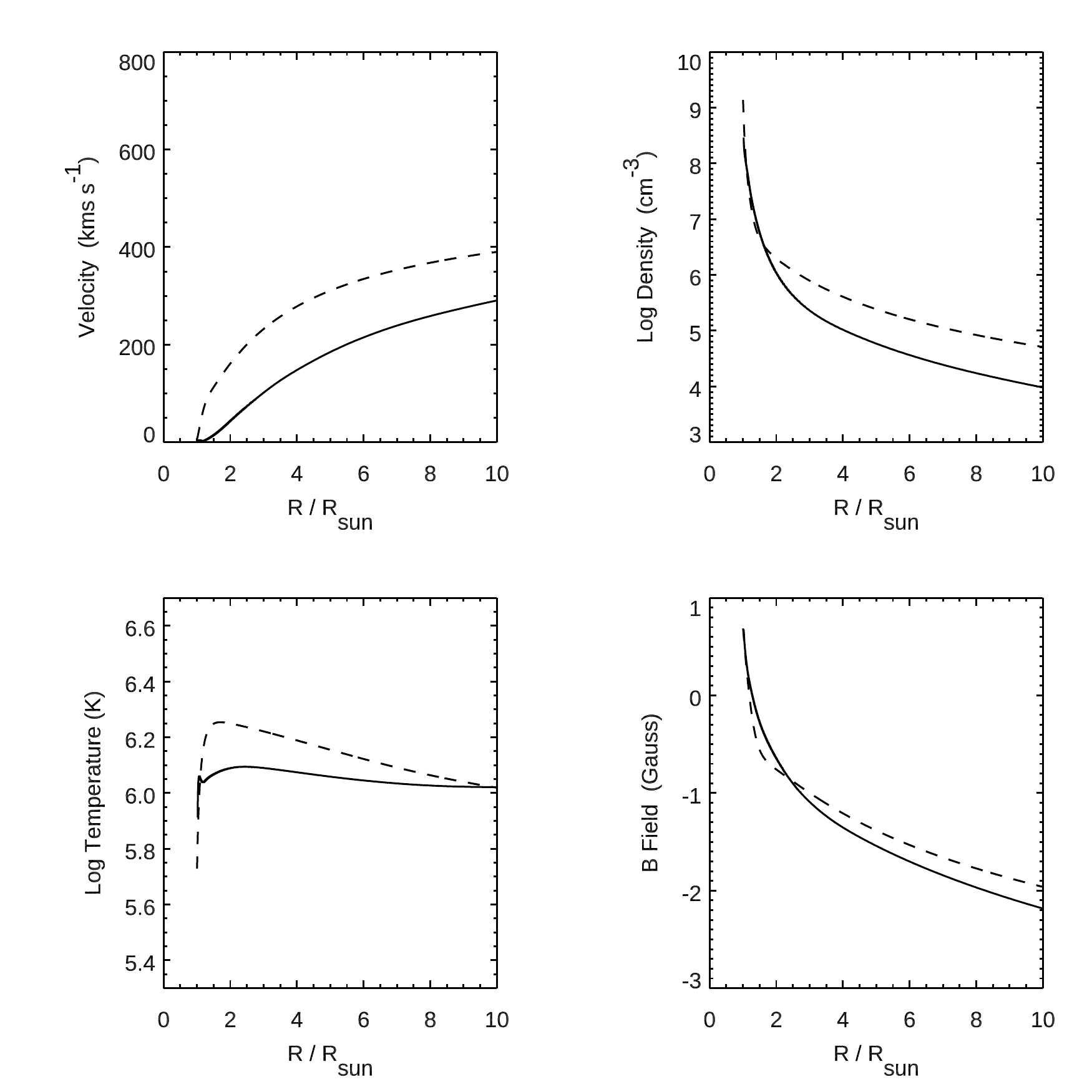}
\caption{Flow parameters along a particular streamline in a pseudostreamer computed with the MAS model, as presented on \citet{shen17} (solid lines).  The dashed lines show a clumpy AWT model similar to that of \citet{asgari-targhi21}, but
for the magnetic field expansion of \citet{shen17} and fractional density inhomogeneities similar to those computed by \citet{asgari-targhi21} (dashed lines).
\label{flow}
}
\end{centering}
\end{figure*}

Figure~\ref{flow} shows two models of the flow along a streamline in a pseudostreamer modeled by \cite{shen17}.  The solid line shows the parameters for one particular field line in the Magnetohydrodynamics on a Sphere (MAS) model \citep{lionello19} that was used by \citep{shen17} to model a pseudostreamer.  In this model, the wind is heated and accelerated by turbulence that is excited when outward propagating Alfv\'{e}n waves interact with Alfv\'{e}n waves reflected higher in the corona.  The dashed line shows the more complex flow predicted by a model like that of \citet{asgari-targhi21} for the fast wind.  It introduces modest density fluctuations and computes the modified rate of Alfv\'{e}n wave reflection and the dissipation heating rate, which in turn affect the flow parameters and temperature.  We will refer to this as the clumpy Alfv\'{e}n Wave Turbulence (AWT) model from now on.  

\begin{figure}
\begin{centering}
\includegraphics[width=3.3in]{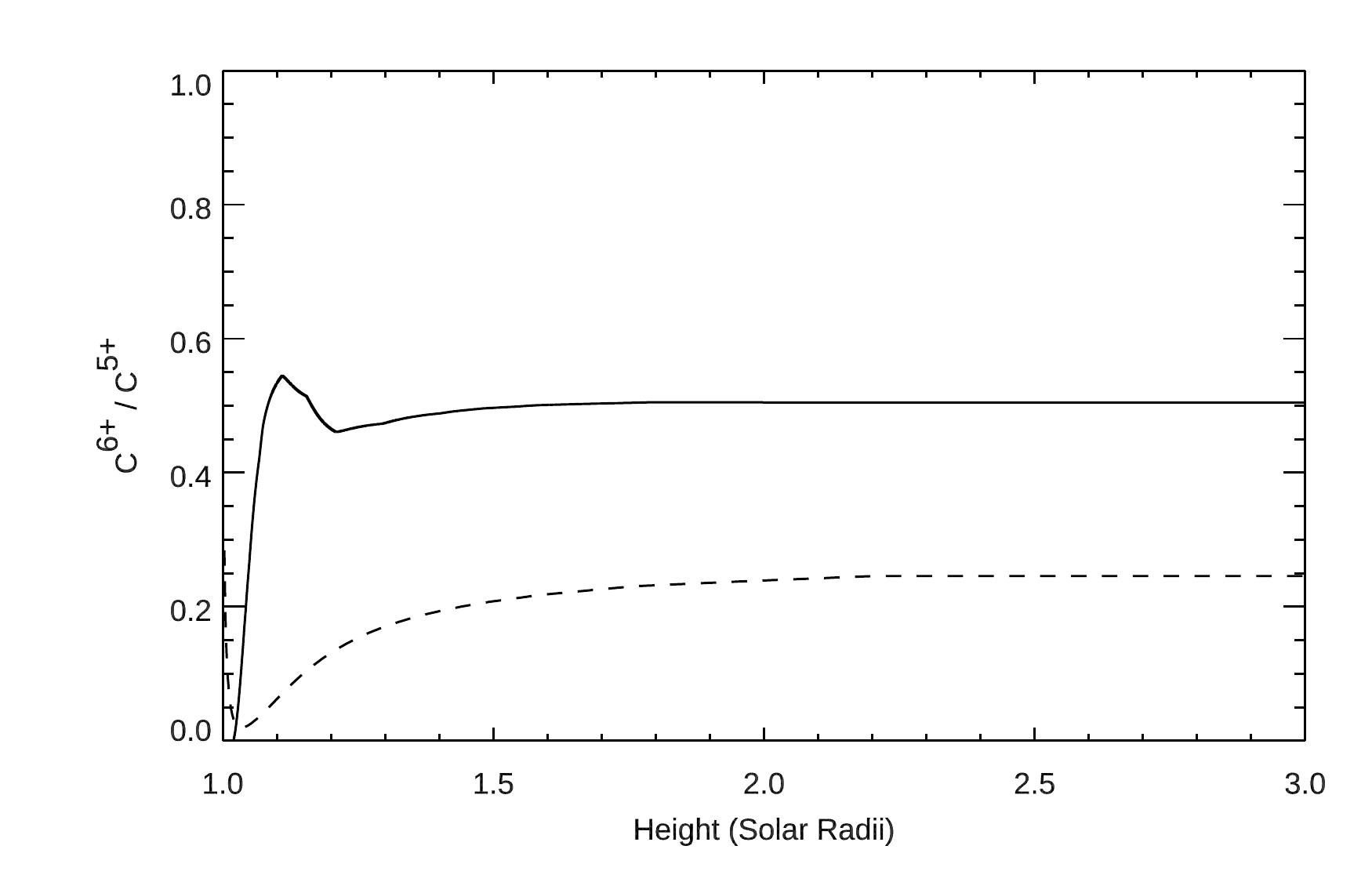}
\caption{\Csix\/ over \Cfive\/ ionization fractions for one streamline of the MAS model (solid line) and the MAS model with the temperature from the clumpy AWT model (dashed line).  The figure emphasizes the large range of possible ionization states, but shows that for this streamline, the \Csix\/ ionization fraction freezes in below 3 \RSUN. 
\label{ionstate}
}
\end{centering}
\end{figure}

Figure~\ref{ionstate} shows the ratios of \Csix\/ and \Cfive\/ ion fractions for two models based on a streamline along the pseudostreamer edge from \citet{shen17}, one using the temperature structure of that model, and one using the clumpy AWT model.   We note that although the temperature in the clumpy AWT model is higher, the rapid acceleration and corresponding low density lead to a lower frozen-in ionization state in this model.   Time-dependent ionization states were computed using ionization and recombination rates from CHIANTI \citep{dere19} and the PlasmaPy-NEI code in plasmapy of \citet{shen17} and \citet{wilson22}, as well as an older code of \citet{raymond79}.  In both models for this streamline, as well as for a dozen others we investigated, the \Csix\/ ionization fraction freezes in around 2 \RSUN, so we will use that value.  We note that there is a longstanding discrepancy between ionization states predicted by solar wind models and charge states measured at 1 AU, in the sense that the models predict lower ionization than are observed \citep{ko96, landi14, oran15, shen17, lionello19, rivera2020}, suggesting that the freeze-in may be slightly higher than indicated by the models.

The freeze-in conditions for ICMEs are much less clear because of their large ranges of densities and outflow speeds, except that continued heating after the material leaves the solar surface is required to match the ionization states observed at 1 AU \citep{rakowski07, rakowski11}.  It is not clear whether the \Csix\/ dropouts originate in prominence ejecta, leading edge material, the flux rope void, or the EUV dimming regions that often accompany the CME eruption.  Prominence material is denser and it expands more slowly than the surrounding material, and it is more likely to have an excess He abundance \citep{delzanna04}, though chromospheric evaporation can also lead to enhanced He abundances \citep{fu20}. 
In the particular CME modeled by \citet{rivera19}, the prominence material \Csix\/ was not yet frozen even at 8 \RSUN, though the hottest component freezes in by 4 \RSUN.  Thus there is great uncertainty as to the freeze-in height even within a single CME, and CMEs show a wide range of velocity and density profiles. We will take 4 \RSUN\/ to be a characteristic freeze-in height for CMEs, but we keep in mind that it will be higher for dense prominence material, especially in slow CMEs.

\begin{figure*}
\begin{centering}
\includegraphics[width=6.0in]{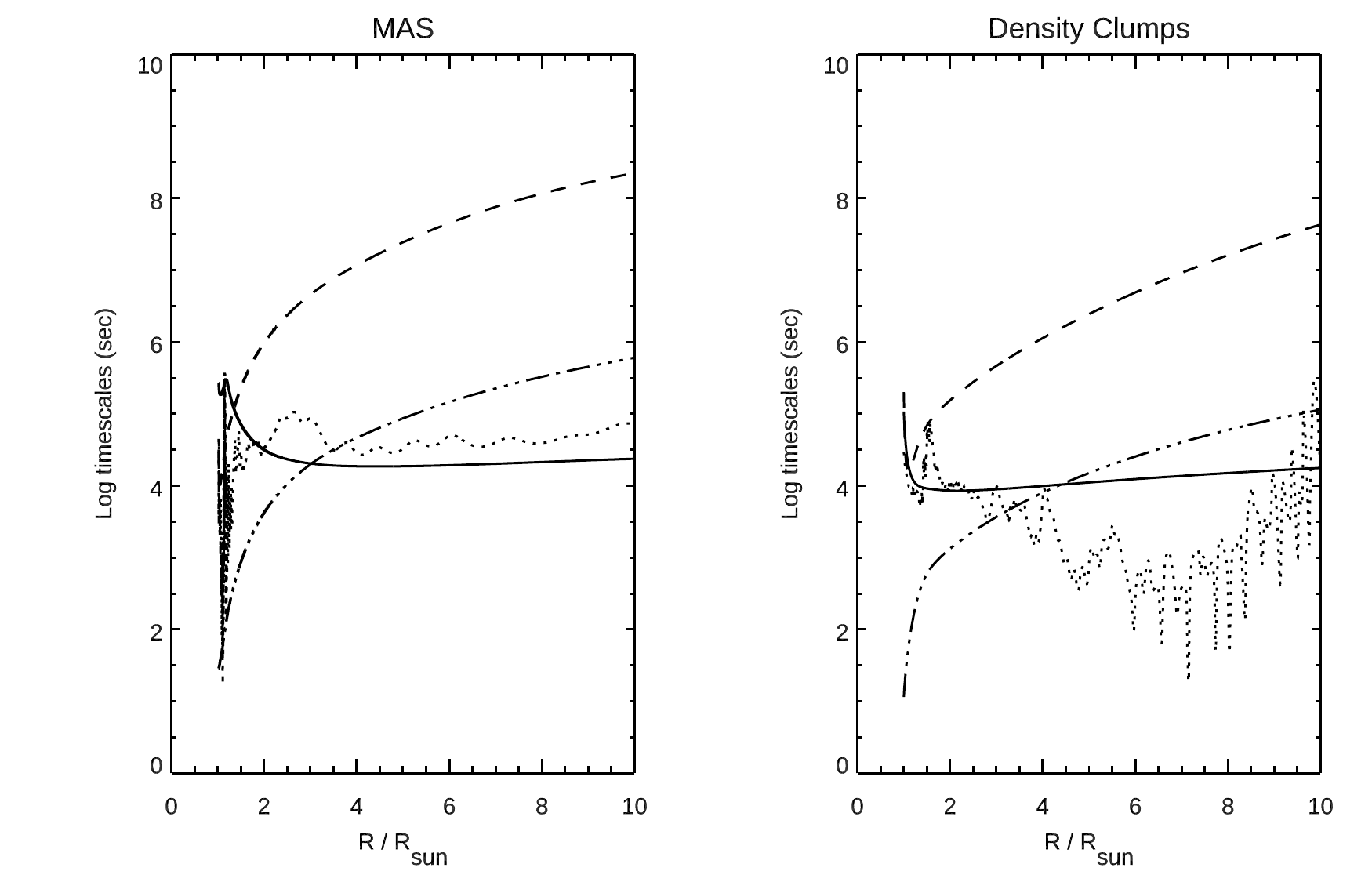}
\caption{Timescales computed for the models shown in Figure~\ref{flow}.  For each model, the flow time (taken to be 0.3 R/V; solid line), ionization timescale for \Cfive\/ (1./(density $\times$ ionization rate); dashed line), Coulomb energy loss timescale (dash-dot) and heating timescale (dotted) are given.  The smooth flow MAS model (\citep{shen17} is shown in the left panel, and the clumpy AWT model in the right panel.
\label{timescales}
}
\end{centering}
\end{figure*}

\subsection{Transition to collisionless heating}

Collisionless plasmas behave differently than collisional ones in many ways. For the purposes of this paper, the relevant condition is that cyclotron resonant heating requires that the collision frequency for an ion to lose energy be less than the cyclotron frequency \citep{hollweg99}. We assume that the Coulomb collisional timescale \citep{spitzer68} is most important.  It depends on the density and temperature of the plasma.  Figure~\ref{timescales} indicates that the plasma becomes collisionless above about 3 \RSUN\/ for this streamline.   As indicated by preferential heating of oxygen, roughly 2.5 to 3 \RSUN\/ is a reasonable estimate for the height of the transition to collisionless conditions in the slow wind \citep{frazin03}.

\subsection{Gravitational Potential}

The fractionation must occur close to the Sun because it is difficult for ions to move very far relative to the wind in the highly supersonic flow at larger radii.  Turbulent heating occurs in ICMEs at 1 AU, but it is modest \citep{liu06}.  In addition, merely shifting the fully stripped ions within the wind would create regions of overabundance of these ions as well as regions of underabundance, and no such regions of overabundances are observed. In order to separate \Csix\/ from the other carbon ions, it is necessary to have both a different heating rate, which we attribute to cyclotron resonant absorption, and another force, such as gravity, which can act on all the ions, but is countered to different degrees by the thermal content of different ions.  Carbon ions are 12 times heavier than protons, but once the wind becomes supersonic, gravity can have only a modest effect on the overall flow.  The fractionation must occur where the solar gravity is still strong enough to separate heavy ions such as \Csix\/ from lighter protons and \Hetwo\/ ions.

A rough criterion for the maximum height of the separation is

\begin{equation}
E_G = E_{TH} + E_{KIN}    
\end{equation}

\noindent
where E$_G$ is the gravitational potential energy, E$_{TH}$ the thermal energy, and E$_{KIN}$ the kinetic energy of the bulk flow.  For carbon ions, these are 4000/R eV, 90 T$_6$ eV, and 600 V$_{100}^2$ eV, respectively, where R is in solar radii, T$_6$ is the temperature in units of MK, and V$_{100}$ is the wind speed in units of 100 \kms.  For the streamline shown in Figure~\ref{flow}, this height is around 8 \RSUN if carbon is not preferentially heated.  This estimate is likely to be altered by other forces, including the ponderomotive force of Alfv\'{e}n waves, drag between carbon ions and protons, and Lorentz forces.  However, it gives a rough upper limit for the height at which the fractionation takes place.  Note that this height is below 3 \RSUN\/ in the fast solar wind because of the rapid acceleration there \citep{cranmer99}.

\section{Wave Dissipation or Reconnection?}

Assuming that the dropouts are caused by deficits in the resonant cyclotron heating rate, they could result from a deficit in wave power due to reduced dissipation of wave energy, to absorption of wave power by excess \Hetwo, or to excess density, beyond the capacity of the wave power to provide enough heat.  These possibilities are not mutually exclusive.  For instance, reduced wave power could cause the \Hetwo\/ outflow to stall, increasing the \Hetwo\/ concentration and further reducing the wave energy available per particle.  

Another model for heating the ions by turbulence generated by dissipation of Alv\'{e}n waves was introduced in response to problems getting the turbulent cascade to reach the relevant cyclotron freequencies.  In this model, the perpendicular cascade produces lower-frequency KAWs that can stochastically heat the ions when they reach the scale of the ion gyroradii \citep{chandran10}.  This mechanism could also starve the fully-stripped ions of heating if \Hetwo\/ or protons absorb too much energy.  Indeed, \citet{chandran10} found that the wave power must exceed a critical value in order for irreversible heating to occur.  A difficulty arises from the relative gyroradii of diffferent ions, however.   The gyroradii scale as m$_i$v/q$_i$, and if the ions are in thermal equilibrium with each other at low heights where the heating must begin to operate, they scale as m$_i^{1/2}$/q$_i$.  The relative gyroradii of H:\Hetwo:\Csix:\Oeight:Si$^{10+}$:Fe$^{11+}$ are thus 1.0:1:0.58:0.5:0.53:0.68.  In other words, \Hetwo\/ and protons could absorb the cascading wave energy before it reached the gyroradii of fully-stripped ions such as \Csix, but they would also prevent the cascade from reaching the gyroradii of most other coronal ions.  It is possible that stochastic heating could operate after some preferential heating has occurred so that the ion velocities are similar to each other rather than proportional to m$_i^{-1/2}$, but it seems likely that the cyclotron resonant heating must dominate the formation of the outliers.

\subsection{Dissipation of Alfv\'{e}nic waves}

A popular mechanism for driving the solar wind is based on Alfv\'{e}n waves propagating up from the chromosphere.  Left to themselves, Alfv\'{e}n waves do not dissipate efficiently in the corona.  However, they can be partially reflected by the gradient in the Alfv\'{e}n speed in the corona, and the reflected waves interact with the outward-going waves to produce turbulence and heating \citep{verdini10, cranmer12}.  That mechanism is the basis of the heating term in two successful MHD models of the global corona and wind; AWSoM and MAS \citep{vanderholst14, lionello19}.  In order for the turbulent energy to be dissipated, it must cascade to small scales.  Much of the energy may dissipate at the proton scale, but some will be absorbed by resonant cyclotron heating of higher mass-to-charge ratio ions before it reaches that scale.

There is observational evidence for wave heating of high m/q ions, though it does not discriminate between resonant cyclotron absorption of MHD waves and stochastic heating.  The line widths of the Ly$\alpha$, O VI and Mg X ultraviolet lines measured in coronal holes by UVCS imply kinetic temperatures of the heavier ions that are more than mass-proportional; $T_O > 16 T_p$ and $T_{Mg} > 24 T_p$ \citep{kohl97, cranmer99}.  This requires strong preferential heating of oxygen and magnesium, which is usually attributed to absorption of wave energy at the cyclotron resonant frequencies as the turbulent cascade transfers energy to small scales and high frequencies \citep{cranmer99, hollweg06}, though it could also be explained by stochastic heating by lower frequency waves \citep{chandran10}.  A cyclotron resonance heating process might also explain the factor of 100 variation in the $^3$He/$^4$He ratio over the course of a Carrington rotation reported by \citet{gloeckler16}. 


While the overall picture is attractive, the Alfv\'{e}n wave reflection in a smoothly accelerating wind is not strong enough to generate adequate dissipation \citep{vanballegooijen16, asgari-targhi21}.  One way to solve the problem is to introduce density fluctuations that locally increase the reflection and dissipation.  Density fluctuations of a few percent to around 20 percent are known to exist in the 2 to 10 \RSUN\/ range \citep{miyamoto14, hahn18, cadavid19}. The reflection produces very localized heating, but the reflection coefficient drops to zero where $dV_A/dr = 0$ \citep{chandran09, cranmer10}, so places where the density gradient equals the gradient in $B^2$ would be places of minimal heating, while the heating would peak where the density increases as B decreases. 

\begin{figure}
\begin{centering}
\includegraphics[width=3.4in]{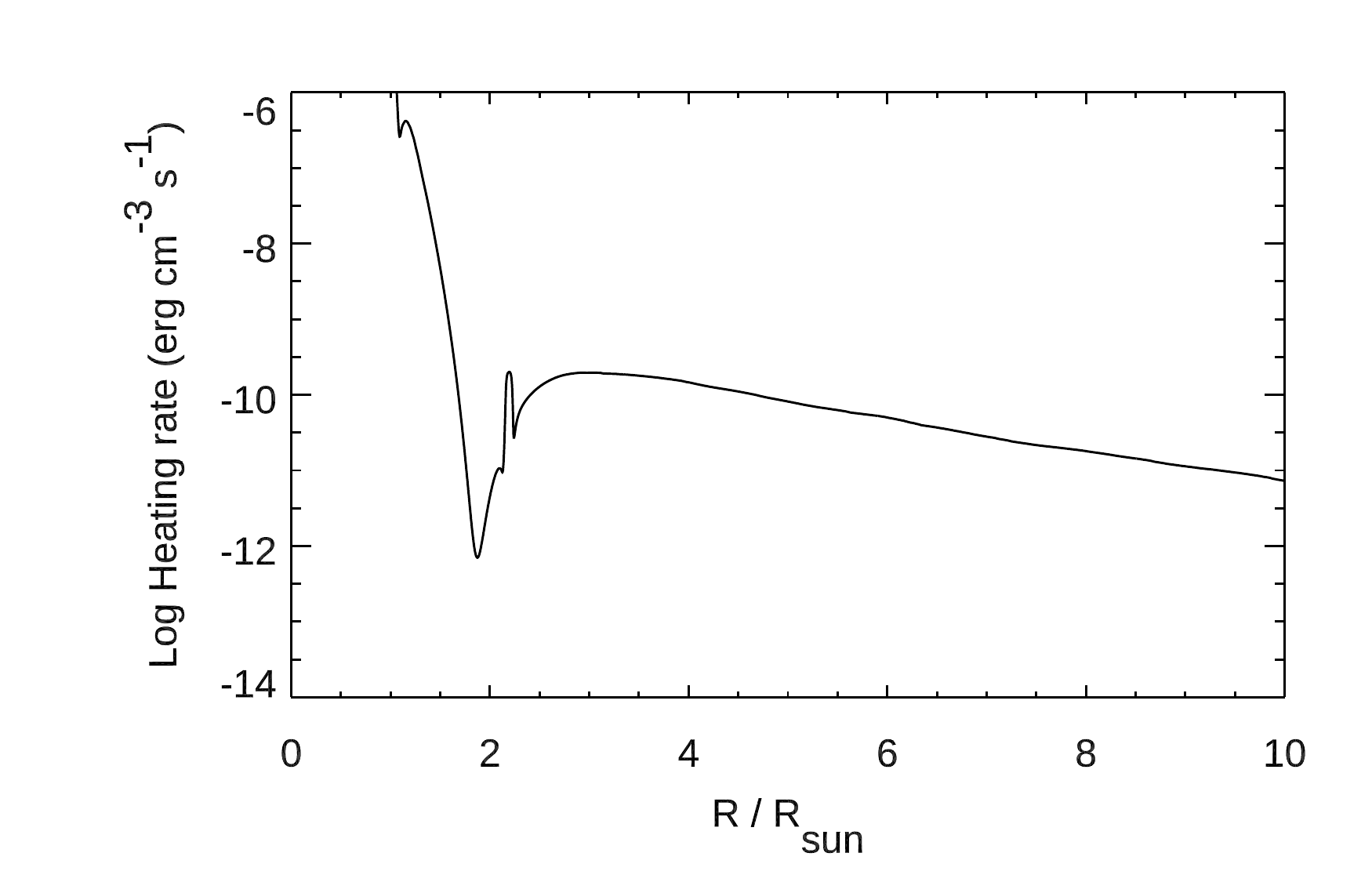}
\caption{Heating rate along the streamline shown in Figure~\ref{flow} for the clumpy AWT model.  Excursions above and below the average by an order of magnitude occur on small scales, especially below 5 \RSUN.
\label{heating}
}
\end{centering}
\end{figure}

Figure~\ref{heating} shows the time-averaged heating rate along the slow wind stream shown in Figure~\ref{flow} as computed by the code of \citet{asgari-targhi21}, with similar modest (24\%) random density fluctuations on scales of 0.05 \RSUN\/ (clumpy AWT model). These Reduced MHD (RMHD) models start with a smooth solution to the 1D flow in a flux tube with heating by Alfv\'{e}n wave dissipation, then add density fluctuations at the observed level, recompute the wave reflection and heating, and obtain self-consistent time-steady solutions. The heating rate in $\rm erg~cm^{-3}~s^{-1}$, that is to say the locally dissipated wave power that drives the turbulent cascade, varies by orders of magnitude.  The local variations on short time scales span two orders of magnitude about the average at the lower heights, though they diminish above about 7 \RSUN.  \citet{chandran10} found that the heating rates corresponding to Alfv\'{e}n wave reflection in a smooth outflow were roughly at the critical levels needed to preferentially heat \Hetwo\/ and \Ofive\/ in the stochastic heating model.  Therefore, the strong heating regions in the clumpy AWT models could easily produce preferential heating, while the weak heating regions could not.  Thus it is possible that the dropouts correspond to places where the heating rate Q is unusually small. In particular, the low heating region around 2 \RSUN\/ in  this particular model is a place where heating of bare ions could be greatly reduced.


As a hypothesis to explain the \Csix\/ dropouts in the slow wind by a lack of resonant cyclotron heating, we suggest that a blob of denser gas from the chromosphere or a prominence, possibly rich in He and Ne, is lofted to the height of ionization freeze-in and the transition from collisional to collisionless conditions.  It could be driven by either MHD or pressure forces, but the former seems more likely.  If it reaches that height but has not reached escape speed, it will fall back unless Alfv\'{e}nic waves can heat and drive it onwards.  Protons are light enough that their thermal energy might still allow them to escape, and most other ions can be heated by cyclotron resonance.  However, if He$^{2+}$ absorbs so much of the energy at its resonant frequency that it falls below the critical value for irreversible heating \citep{chandran10}, or if the heating rate is low because of the wave reflection properties of the blob, then the heavier bare ions will be unable to overcome gravity, and they will be left behind.  \citet{chandran10} indicate that level of turbulence in the steady-flow Alfv\'{e}n wave reflection models lies above the critical value for resonant heating for plausible parameter choices, but the low points in the heating rate shown in Fig.~\ref{heating} would not.

\subsection{Reconnection}

The idea that the dropouts are associated with reconnection is supported by the magnetic field signatures similar to SIMFRs seen in dropout events and by the frequent occurrence of dropouts in ICMEs \citep{rivera21}.  Reconnection plays a dominant role in solar flares, and it is observed at larger heights as downflows and disconnection events (bifurcated flows) in CME current sheets seen in white light \citep{savage10} and especially near sector boundaries in quiet Sun streamers \citep{wang99b,sheeley14}.  The FIP effect in the slow solar wind is generally attributed to the release of material from closed loops onto open field lines by reconnection \citep{schwadron99, laming17}.  Reconnection is also a fundamental part of the S-web picture of pseudo-streamer winds \citep{higginson17}, though it may occur at heights well below the freeze-in height.  \citet{scott22} have explored the ionization state signatures of interchange reconnection at various heights.  They predict regions of enhanced \Oseven/\Osix, but there is no reason to expect \Csix\/ dropouts from time-dependent ionization in those regions.

However, there is good reason to expect that very strong wave turbulence is present in reconnection regions.  Excess widths of spectral lines in the current sheets that stretch between flare loop tops and CME flux ropes are interpreted as turbulent velocities of order 100 \kms \citep{ciaravella08, bemporad08, warren18}, and velocity variations are also observed \citep{freed18}.  Cyclotron resonance interaction is believed to be responsible for selective acceleration of $^3$He \citep{fisk78} in flares, and perhaps $^{22}$Ne \citep{mewaldt79}. Moreover, turbulence is central to the acceleration process that produces the power-law distributions of electrons seen in solar flares.  On the theoretical side, turbulence is produced by tearing modes or other instabilities, and it is crucial to rapid reconnection \citep{lazarian99, loureiro12, shen13}.   Turbulence within the current sheet may generate waves that propagate into surrounding plasma which does not pass through the current sheet itself, but it is not clear how much turbulent energy would be available to this less strongly heated and ionized gas \citep{ye21}.  

Overall, it seems unlikely that a current sheet is a promising location where a lack of wave power at the bare ion cyclotron frequency could cause dropouts.  However, it is entirely plausible that reconnection events inject high density, high He abundance plasma into regions where the ionization state has frozen-in, and where fractionation can occur.

\section{Slow Wind, Fast Wind and ICMEs} \label{sec:slowfast}

We now ask how the Alfv\'{e}nic wave scenario would play out in the fast and slow solar winds and in ICMEs.

\subsection{Fast Solar Wind}

Since almost no Outlier events are observed in the fast solar wind, the question becomes why the mechanism that fractionates the ions does not operate there.    \citet{cadavid19} and \citet{hahn18} showed that density fluctuations are present, and \citet{asgari-targhi21} showed that those density fluctuations produce very large variations in the heating rate per particle.  Moreover, polar jets have been observed with EIT, UVCS and LASCO \citep{gurman98, dobrzycka02, wang98, uritsky20}. Perhaps the wave power is so large that even near minima in the heating rate, the power at the cyclotron resonance remains above the critical value described by \citet{chandran10}.  This would be compatible with the rapid acceleration of the wind in coronal holes.  Another possibility is that the freeze-in radius could be near or above the sonic point, so that reduced heating does not mean that the \Csix\/ ions fail to escape.  As mentioned above, the rapid acceleration of the fast wind \citep{cranmer99} means that $\rm E_g = E_{KIN} + E_{TH}$ below 3 \RSUN, and dropouts may not be able to develop.

\subsection{Slow Solar Wind}

\citet{cadavid19} showed that density fluctuations are present in the quiet corona as well as in coronal holes, and the quiet Sun and active regions contain high coronal loops that could release plasma into the slow solar wind. We have shown that the predicted fluctuations in the turbulent heating rate are also large, particularly in the region around 3 \RSUN\/ where the ionization freeze-in occurs.  This is the region where resonant heating is particularly important, in that \citet{frazin03} found that the thermal velocities of O$^{5+}$ gradually approach those of protons between 2.5 and 5 \RSUN\/ in streamers and that the O$^{5+}$ velocities seem to become anisotropic above about 4 \RSUN.  Models for the FIP effect that involve release of plasma from closed loops would naturally provide the required density enhancements provided that the loops are high or that the plasma is released at high enough speed to reach 3 to 4 \RSUN. The increase in magnetic fluctuations downstream of the outliers could arise naturally from the reflection of Alfv\'{e}nic waves traveling outwards from the Sun.

Therefore the wave heating scenario suggested above, in which denser blobs of gas are lifted to the freeze-in region, and all but the bare ions are heated and driven by turbulence,  would also fit in with the observed density enhancements and the lower temperatures measured in the Outliers.  Thus the dissipation of Alfv\'{e}nic waves seems well-suited as an explanation for the \Csix\/ dropouts in the slow wind.  The main feature of the Outliers that is not explained easily in this picture is the rotation in magnetic field direction.  That might be explained if the blobs were launched by reconnection events and the magnetic structure was preserved in the fractionation region.

An alternative picture for the origin of density enhancements is suggested by \citet{hahn22}.  They propose that a parametric decay instability, in which a high amplitude outward-propagating Alfv\'{e}n wave decays to an inward-propagating Alfv\'{e}n wave and an outward-propagating acoustic wave, can account for the density and velocity patterns they observe in the Si IV lines with IRIS.  Their observations pertain to the transition region at lower heights, so it is not clear whether this instability is important where the ionization state freezes in.  It is also unclear how the process could produce the magnetic and abundance signatures seen in the dropouts.  Nevertheless, it should produce density fluctuations of sufficient magnitude to drastically alter the heating rate, and therefore the turbulent energy available to heat the ions.

The reconnection scenario was invoked by \citet{rivera21}.  Blobs of gas that detach from the cusps of streamers are believed to be created by reconnection, then to accelerate up to slow solar wind speed \citep{wang99a, sheeley97}.  However, these LASCO blobs have been associated with the heavy ion dropouts detected at 1 AU by ACE \citep{weberg12, weberg15} and interpreted as the release of material from the closed loops of the streamer core that had undergone gravitational settling \citep{raymond97}.  No such elemental depletion is seen in the Outliers.

\citet{brooks20} invoke a two-component origin for the slow solar wind based on Hi-C and AIA observations of an active region.  They suggest that reconnection between closed loops and open field lines injects FIP-enhanced material, while the flow emerging from the active region plage has photospheric abundances. The component injected by reconnection would perhaps be subject to the ionic fractionation that produces Outliers.  \citet{scott22} predict regions of enhanced density just below regions of enhanced \Oseven/\Osix, and the high density regions could reflect upward-propagating Alfv\'{e}n waves before they reach the high ionization zone, thus starving that region of wave heating and producing \Csix\/ dropouts. The large percentage of the slow solar wind that shows \Csix\/ depletion, around 10\%, would imply that a significant fraction of this component has the right conditions above the freeze-in height to become fractionated.

\subsection{ICMEs}

In the common picture of CME eruptions, a flux rope becomes unstable and is launched by MHD forces.  Magnetic reconnection accompanies the ejection, both as a cause and an effect of the eruption, and some of the reconnecting field lines form a flux rope or a sheath around an existing flux rope \citep{gosling95, lin04}.  The fractionation of bare ions indicates that cyclotron resonant waves play a role, so that MHD forces are not the only processes involved.  As mentioned above, there is evidence from UV and EUV spectra and from ionization states measured at 1 AU that the erupting gas experiences continued heating at least up to several \RSUN.  The nature of that heating has not been firmly established, but relaxation of the expanding, stressed magnetic field toward its minimum energy state \citep{lynch04} requires the release of energy, most likely by way of reconnection.  

On the other hand, the EUV dimmings observed in many CMEs \citep{dissauer18} indicate that a large amount of material is launched as magnetic loops straighten toward radial, and density inhomogeneities would be subject to the Alfv\'{e}nic wave mechanism described above. In some cases, material launched in the eruption is observed to fall back to the solar surface when it is unable to escape \citep{innes16}.  Thus wave-driven fractionation is also possible.

Several pieces of evidence may favor one mechanism or the other.  The magnetic structure, electron strahl and overabundance of He and Ne in the Outliers could be explained if the material originates in prominence flux ropes, since there are observations of high He and Ne abundances in prominences \citep{spicer98, delzanna04, li20}.  Prominence material is also indicated by low charge states, such as singly ionized He \citep{yao10}.  On the other hand, to the extent that plasma is trapped in the dips in a magnetic flux rope, it is difficult for fractionation to occur, though flows along flux ropes can occur as the flux ropes straighten out.  The generally high ionization states seen in ICME Outliers \citep{kocher17} suggest that much of the material experienced heating in a reconnection current sheet.  The EUV dimmings can account for a large amount of mass, but it is uncertain how much that material is heated and what ionization state would be expected.  Also, the material might lie outside the flux rope, and therefore not show the observed magnetic structure, though it could show electron strahl.  A large percentage of ICMEs show Outliers, 72\%, implying a substantial filling factor.  That could be consistent with gas from the EUV dimming regions.  A large filling factor would also be consistent with plasma from the main reconnection current sheet, which fills the flux rope formed by reconnection itself.  However, a very large expansion factor between the current sheet and the outer flux rope might lead to low densities, in contradiction to the observations.

\section{Conclusions}

\citet{zhao17, kocher17} discovered periods in the slow solar wind and ICMEs when the bare ions of carbon and other elements were severely depleted.  The fact that the fully stripped ions of C, N and O all show drastic depletions \citep{zhao17, kocher17} strongly suggests that the cyclotron frequency that they share with \Hetwo is a key factor and that the strong preferential heating of heavier ions \citep{kohl97} is weakened, while the magnetic field anomalies similar to SIMFRs \citep{rivera21} strongly suggest that reconnection plays a significant role.  We therefore hypothesize that reconnection injects cloudlets of relatively dense, perhaps He-rich plasma into the corona, and that such cloudlets in the region above ionization freeze-in and the collisional-collisionless transition, but below the height where all the ions flow freely to 1 AU are the origin of the dropouts.  It is also possible that some other mechanism, such as parametric decay instabilities \citep{hahn22} produces the density fluctuations.

A plausible, but unproven, interpretation is that \Hetwo\/ absorbs so much of the resonant wave power that the heating rate of the bare ions falls below a critical value for preferential heating \citep{chandran10}, and that the bare carbon and heavier ions are unable to escape the Sun's gravitational potential.  Density fluctuations and the resulting changes in Alfv\'{e}n speed would lead to large variations in Alfv\'{e}n wave reflection and turbulent heating rates \citep{asgari-targhi21}.  The clumpiness of the solar wind provides some support for this idea.  The analysis by \citet{brooks20} indicating that the slow wind contains both material from opening active region loops and material injected from the chromosphere is consistent with the general picture, though the magnitudes of the density contrasts, the sizes of the clumps, and the heights where the clumps occur are not well known.

Thus we suggest that density inhomogeneities between roughly 3 and 7 \RSUN\/ are created by reconnection in the corona, chromosphere or prominences, and that they produce regions where weakened preferential heating of bare ions is unable to overcome gravity, and the bare ions fail to escape from the Sun.  The rapid acceleration of the fast solar wind might mean that region where such weakened heating would prevent the escape is too close to the solar surface for the mechanism to operate effectively.  ICMEs have such a broad range of temperatures, heating rates and accelerations that it is difficult to pin down where such a mechanism would be most effective, but it is known that heating continues after the initial ejection from the solar surface \citep{rakowski07, murphy11, wilson22}.

If our hypothesis that the common cyclotron frequency of \Hetwo\/ and the other fully stripped ions accounts for the dropouts is correct, it supports the resonant cyclotron absorption models \citep{hollweg02} for coronal heating over the models that invoke stochastic heating by lower frequency waves \citep{chandran10}.  It is therefore important to test this hypothesis further.  It is difficult to test with remote sensing observations, because the bare ions have no spectral lines, and the dropouts probably form well below the regions probed by PSP or Solar Orbiter.  Therefore, those tests are likely to be indirect or model-dependent. 

We are far from a complete theory of the origin of the bare ion dropout events, but we have attempted to constrain the region where they originate.  In particular, the observed density fluctuations are not quantitatively understood, there are no secure numbers for the wave power spectrum, and we have not considered the ponderomotive force or ion-ion drag.   Transverse density variations \citep{downs13, raymond14} may severely affect the Alfv\'{e}n wave reflection computed from 1D models.  Nevertheless, if the general picture is correct, we might expect that Parker Solar Probe and Solar Orbiter might measure density enhancements, He abundance enhancements and magnetic signatures of flux ropes with timescales on the order of an hour in the radial direction, though the transverse size scales are unknown.  If fully-stripped ions are detected, they might show lower velocities.








\acknowledgments
This work was carried out as part of NASA Grant 80NSSC19K0853 to the Smithsonian Astrophysical Observatory. MAT is supported under contract 80NSSC18K1207 from the
NASA Heliophysics Supporting Research program to the
Smithsonian Astrophysical Observatory (SAO) and Columbia University. YJR acknowledges support from the Future Faculty Leaders postdoctoral fellowship at Harvard University. STL acknowledges support from NASA Grants 80NSSC19K0853, 80NSSC20K0185, 80NSSC18K0645,80NSSC22K0204, and 80NSSC20K0192 to the University of Michigan.   {\bf We thank the referee, Martin Laming, for valuable insights that have strengthened the paper.}

%

\vspace{5mm}


\software{plasmapy}




\bibliography{export-bibtex.bib}
\bibliographystyle{aasjournal}





\end{document}